\documentstyle[psfig]{caosp}
\begin{document}
\htitle{Is Vega a $\lambda$ Boo star?}
\hauthor{S. Iliji{\'{c}} {\it et al.}}
\title{An abundance analysis for Vega: Is it a $\lambda$ Boo star?}
\author{S. Iliji{\'{c}} \inst{1} \and M. Rosandi{\'{c}} \inst{1} \and
D. Dominis \inst{1} \and M. Planini\'{c} \inst{2} \and
 K. Pavlovski \inst{1}}
\institute{University of Zagreb, Faculty of Geodesy, Ul. Andrije Ka\v{c}i\'{c}a
Mio\v{s}i\'{c}a 26, 10000 Zagreb, Croatia \and University of Zagreb, Faculty of
Sciences, Bijeni\v{c}ka 32, 10000 Zagreb, Croatia}

\maketitle
\begin{abstract}
Since Baschek \& Slettebak (1988) drew attention to the similarity between
the abundance pattern of $\lambda$ Boo stars and that of Vega, there has been
a long debate whether Vega should be listed among the chemically peculiar stars 
of $\lambda$ Boo type. We performed an elemental abundance
analysis using a high dispersion spectrum in the optical region, and confirmed 
its mild metal underabundance. In our discussion we reinforce the suggestion 
that Vega is a mild $\lambda$ Boo star.
\end{abstract}
\section{Introduction}

Vega ($\alpha$ Lyr = HD 172167 = HR 7001) is a Population I star of
spectral type A0 V, with a projected rotational velocity $v \sin i =
23$ km$\,$s$^{-1}$. It has been extensively studied both for its role
of primary spectrophotometric standard in the visual and in the UV,
respectively, and for its role of comparison star in abundance studies
of A-type stars.

Vega has a distinctly non--solar composition. Since Baschek \& Slettebak
(1988) drew attention to the similarity between the abundance pattern of 
$\lambda$ Boo stars and that of Vega, there has been a long debate whether Vega 
should be listed among the chemically peculiar stars of $\lambda$ Boo type.

Moreover, Baschek \& Slettebak suggested that the $\lambda$ Boo stars
may be regarded as {\em rotating Vegas}, or conversely Vega may be
regarded as a {\em mild non--rotating $\lambda$ Boo star}.

\section{An abundance analysis}

High dispersion spectrum (4 {\AA}$\,$mm$^{-1}$) of Vega was obtained with
the Coud\'{e} spectrograph of the Ond\v{r}ejov Observatory 2--m telescope
on 19 August 1992. The blue region was covered, $\lambda\lambda$ 4043 to
5057 {\AA}.

The abundances of individual elements are derived by the curve-of-growth method.
Altogether, 245 lines of 16 elements have been identified and their equivalent
widths were measured. The mean microturbulent velocity was $v_t = 1.75$
km$\,$s$^{-1}$.

\begin{table}[t]
\small
\begin{center}
\caption{Abundances derived for Vega in the present work (the number of
lines used is indicated in parentheses), and compared with the
study by Adelman \& Gulliver (1990, hereafter AG),
and solar composition as derived by Anders \& Grevesse (1989).}
\begin{tabular}{lccc|lccc}
\hline\hline
Element &  AG & This work & Sun     & Element&AG    & This work   & Sun     \\
\hline
Cr I &      & -6.04 (3)  & -6.33  &   Mg II&-5.11  & -5.05 (3)  & -4.42   \\
Cr II&-6.76 & -6.52 (2)  & -6.33  &   Si II&       & -4.85 (2)  & -4.49   \\
Mn I &-7.16 & -7.22 (2)  & -6.61  &   Ca I &-6.21  & -6.07 (2)  & -5.64   \\
Fe I &-5.05 & -5.29 (4)  & -4.33  &   Sc II&-9.62  & -9.46 (3)  & -8.90   \\
Fe II&-5.12 & -4.73 (4)  & -4.33  &   Ti I &       & -7.51 (2)  & -7.01   \\
Ni I &-6.38 & -6.60 (2)  & -5.75  &   Ti II&-7.47  & -7.19 (7)  & -7.01   \\
Sr II&      & -9.27 (2)  & -9.14  &   V II &       & -7.44 (1)  & -8.04   \\
Zr II&      & -9.08 (2)  & -9.54  &        &    &       &       \\
\hline\hline
\end{tabular}
\end{center}
\end{table}

The results of the abundance analysis are given in Table 1 along with
the results of a previous study by Adelman \& Gulliver (1990).
For reference we also give the solar composition
from Anders \& Grevesse (1989). Abundances are expressed as
$\log \epsilon =\log(N_{\rm el}/N_{\rm tot})$.

\begin{figure}[t]
\centerline{                                                           
\psfig{figure=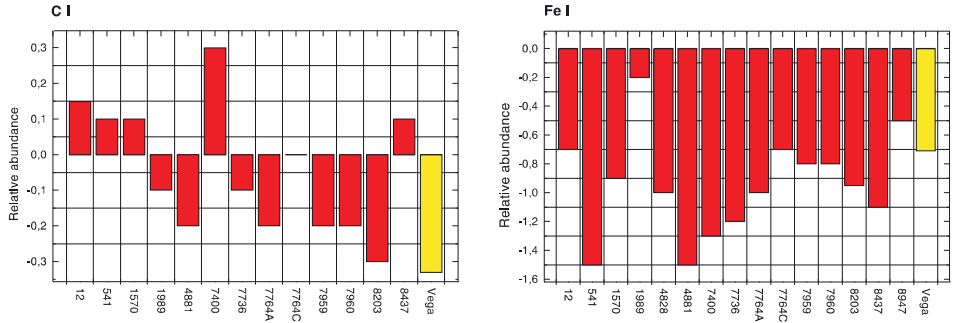,height=4.5cm,clip=}}                                        
\caption{Abundance of carbon (left panel) and iron in Vega compared to abundance
 pattern of $\lambda$ Boo stars as derived by St\"{u}renburg (1993).}                                                 
\end{figure}                                                                

\section{Is Vega a $\lambda$ Boo star?}

The $\lambda$ Boo stars are a class of metal-poor Population I A-type
stars. Since the prototype was recognized by W.W. Morgan in 1943, somewhat
different criteria of classification of $\lambda$ Boo stars were used
(c.f.\ review by Paunzen in the present volume for more details). The most
detailed abundance study of $\lambda$ Boo stars was performed by St\"{u}renburg
(1993) who analyzed 13 stars. The abundance pattern known from previous
studies was confirmed: (i) the light elements (C, N, O and S) have a solar
abundance, and (ii) the heavier elements (Mg, Al, Ca, Fe,...) are
underabundant by up to a factor of 100.

Vega is not included in the recent `Consolidated catalogue of $\lambda$
Bootis stars' by Paunzen et al. (1997) since at classification resolution
it does not share the properties of $\lambda$ Boo stars (c.f. Gray, 1996, for
details). However, its abundance pattern clearly resembles those of $\lambda$
Boo stars as inferred from comparison with St\"{u}renburg's results for his
sample of $\lambda$ Boo stars (Fig. 1). It certainly is slightly metal-weak, and
either 1) may possibly represent what a $\lambda$ Boo star would look like some
tens of millions of years after the {\em $\lambda$ Boo mechanism} turns off,
or 2) may derive its properties from a similar mechanism to the $\lambda$ Boo
mechanism -- which, presumably, is the accretion of metal-depleted gas --
under slightly different conditions, which would lead to a result similar to,
but not identical with what we see in the $\lambda$ Boo stars
(Gray 1997, priv.\ commun.).
Holweger \& Rentsch--Holm (1995) have found that if Vega was observed at a
more typical inclination, it would follow the correlation between Ca deficiency
and rotation shown by other stars of the $\lambda$ Boo class. In that sense,
Vega is exceptional among so-called Vega-type dusty A stars.
Evidence for rapid rotation of Vega came from luminosity excess (Gray 1988)
and detailed line profile calculations (Gulliver et al. 1994).
The idea that Vega is a rapid rotator seen pole-on
may have an important role in interpreting the observed properties of Vega.

\acknowledgements
The authors are thankful to Dr.~Richard Gray for stimulating correspondence.
KP thanks the staff of the Stellar Dept.\ of the Astronomical Institute of
the Academy of Sciences of the Czech Republic for their help and hospitality
during an observing run in August 1992.
This work has been supported by a MZT Project 007002 of Croatian Ministery
of Science and Technology.

\end{document}